\documentclass[11pt,aps,showpacs,nofootinbib,prd]{revtex4}

\usepackage{graphicx}
\usepackage{amsmath}
\usepackage{amssymb}
\usepackage{bm}
\renewcommand{\baselinestretch}{1.2}
\newcommand{\be}{\begin{equation}}
\newcommand{\ee}{\end{equation}}
\newcommand{\bea}{\begin{eqnarray}}
\newcommand{\eea}{\end{eqnarray}}
\def\br{\hskip-2.5mm/}

\newcommand{\rem}[1]{}
\newsavebox{\PSLASH}
 \sbox{\PSLASH}{$p$\hspace{-1.8mm}/}
 
\renewcommand{\theequation}{\thesection.\arabic{equation}}
\newcounter{saveeqn}
\newcommand{\add}{\addtocounter{equation}{1}}
\newcommand{\alpheqn}{\setcounter{saveeqn}{\value{equation}}%
\setcounter{equation}{0}%
\renewcommand{\theequation}{\mbox{\thesection.\arabic{saveeqn}{\alph{equation}}}}}
\newcommand{\reseteqn}{\setcounter{equation}{\value{saveeqn}}%
\renewcommand{\theequation}{\thesection.\arabic{equation}}}
\newenvironment{nedalph}{\add\alpheqn\begin{eqnarray}}{\end{eqnarray}\reseteqn}
 \newsavebox{\notrightarrow}
 \sbox{\notrightarrow}{$\to$\hspace{-4mm}/}
 
 \newsavebox{\PARTIALSLASH}
 \sbox{\PARTIALSLASH}{$\partial$\hspace{-1.6mm}/}
 
 \newsavebox{\ASLASH}
 \sbox{\ASLASH}{$A$\hspace{-2.1mm}/}
 
 \newsavebox{\KSLASH}
 \sbox{\KSLASH}{$k$\hspace{-1.8mm}/}
 
 \newsavebox{\LSLASH}
 \sbox{\LSLASH}{$\ell$\hspace{-1.8mm}/}
 
 \newsavebox{\QSLASH}
 \sbox{\QSLASH}{$q$\hspace{-1.8mm}/}
 
 \newsavebox{\DSLASH}
 \sbox{\DSLASH}{$D$\hspace{-2.2mm}/}
 
 \newsavebox{\DbfSLASH}
 \sbox{\DbfSLASH}{${\mathbf D}$\hspace{-2.8mm}/}
 
 \newsavebox{\DELVECRIGHT}
 \sbox{\DELVECRIGHT}{$\stackrel{\rightarrow}{\partial}$}
 
 \newcommand{\blue}{\IfColor{\textCadetBlue}{}}
\newcommand{\black}{\IfColor{\textBlack}{}}
\newcommand{\red}{\IfColor{\textRed}{}}
\newcommand{\green}{\IfColor{\textOliveGreen}{}}
\newcommand{\lila}{\IfColor{\textRedViolet}{}}

\begin{document}

\begin{flushright}
IPM/P-2006/049\\ hep-th/0608112\\
{\small{Revised: October 5, 2006}}
\end{flushright}
\title{Axial anomaly of QED in a strong magnetic field and noncommutative anomaly}
\author{N. Sadooghi$^{1,2,}$}\email{sadooghi@sharif.edu}
\author{A. Jafari Salim$^{1}$}
\email{jafari_amir@mehr.sharif.edu} \affiliation{$^{1}$Department of
Physics, Sharif University of
Technology, P.O. Box 11365-9161, Tehran-Iran\\
$^{2}$Institute for Studies in Theoretical Physics and Mathematics
(IPM)\\ School of Physics, P.O. Box 19395-5531, Tehran-Iran}
\date{August 16, 2006}
\begin{abstract}
\noindent The Adler-Bell-Jackiw (ABJ) anomaly of a $3+1$ dimensional
QED is calculated in the presence of a strong magnetic field. It is
shown that in the regime with the lowest Landau level (LLL)
dominance a dimensional reduction from $D=4$ to $D=2$ dimensions
occurs in the longitudinal sector of the low energy effective field
theory. In the chiral limit, the resulting anomaly is therefore
comparable with the axial anomaly of a two dimensional massless
Schwinger model. It is further shown that the $U_{A}(1)$ anomaly of
QED in a strong magnetic field is closely related to the {\it
nonplanar} axial anomaly of a conventional noncommutative $U(1)$
gauge theory.
\end{abstract}
 \maketitle
\section{Introduction}
\subsection{Motivation}
\noindent In the early days of current algebra, before the
development of QCD, it was realized both in model field theories
such as linear sigma model of the baryons and in QED that the
flavor-singlet axial current's conservation is broken by quantum
fluctuations, the famous Adler-Bell-Jackiw (ABJ) triangle anomaly
\cite{adler}, $
\langle\partial_{\mu}j_{5}^{\mu}\rangle=-\frac{g^{2}}{16\pi^{2}}\
F_{\mu\nu}\tilde{F}^{\mu\nu}$ (for recent reviews of $U(1)$ axial
anomaly, see \cite{adler-new}). Later on, it was shown that in QCD
the axial-flavor symmetry, although it is a perfect classical
symmetry of massless quarks, is broken by quantum effects. This
symmetry and its corresponding anomaly are sensitive to the strong
coupling, topological excitations in QCD, and also play an important
role in the properties of the theory's vacuum. The far reaching
consequences of the discovery of the quantum anomalies, in general,
include quantitative predictions of physical amplitudes from anomaly
in global symmetries such as in two photons decay of pions, and
restriction of consistent gauge theory models of particle physics
from cancellation of anomalies in local symmetries such as in
electroweak theory. In studying various field theories, it is
therefore important to calculate the anomalies of their various
global and local symmetries.
\par
In this paper, we will derive the axial anomaly of $3+1$ dimensional
QED in the presence of a strong magnetic field, where an
approximation to the regime of lowest Landau level (LLL) dominance
is justified. Our result includes two main observations, which will
be worked out in this paper:
\par
The first observation is that our resulting $U_{A}(1)$ anomaly in
the presence of a strong magnetic background field is comparable, as
expected, with the axial anomaly of an ordinary $1+1$ dimensional
massless Schwinger model \cite{schwinger}. This is indeed related to
the dimensional reduction $D\to D-2$ in the dynamics of fermion
pairing in a magnetic field, which causes a generation of a fermion
dynamical mass even at the weakest attractive interaction between
fermion in the regime of LLL dominance  \cite{gusynin-1}.
\par
The second, and probably the more interesting observation is the
connection of the $U_{A}(1)$ anomaly of QED in the LLL regime with
the axial anomaly of a conventional noncommutative $U(1)$ gauge
theory (see \cite{noncommutative} for a review of noncommutative
field theory (NCFT), and \cite{ned-4} for a recent review of
noncommutative anomalies). The connection between the dynamics in
relativistic field theories in a strong homogeneous magnetic field
and that in NCFT has been recently studied in \cite{miransky-0}. In
particular, it is shown that the effective action of QED in the LLL
approximation is closely connected to the dynamics of a {\it
modified} noncommutative QED, in which the UV/IR mixing
\cite{minwalla} is absent -- a phenomenon which is also observed in
the Nambu-Jona-Lasinio (NJL) model \cite{miransky-0} and in the
scalar $O(N)$ model \cite{ned-5} in the presence of a strong
magnetic field. The UV/IR mixing in the ordinary noncommutative
field theory manifests itself in the singularity of field theory
amplitudes in two limits of small noncommutativity parameter
$\Theta$ and large UV cutoff $M$ of the theory. As it is argued in
\cite{miransky-0}, the reason for the absence of UV/IR mixing in the
modified noncommutative field theories is the appearance of a
dynamical form factor
$\sim\exp\left(-\mathbf{q}_{\perp}^{2}/4|eB|\right)$ for the photon
fields in the regime of LLL dominance.
\par
Due to this connection between the modified noncommutative field
theory and the ordinary one, it is important to calculate the
Adler-Bell-Jackiw (ABJ) anomaly of modified noncommutative QED and
to compare it with the anomaly of an ordinary noncommutative $U(1)$
gauge theory. As we will show later, the $U_{A}(1)$ anomaly of QED
in the presence of a strong magnetic field is in particular
comparable with the {\it nonplanar} anomaly of a conventional
noncommutative QED (see the second part of this section for a review
of anomalies in noncommutative QED and more detailed comparison).
\par
The organization of this paper is as follows: As next, we will
summarize our results by presenting some necessary technical details
on the relation between the $U_{A}(1)$ anomaly of a $3+1$
dimensional QED in the presence of a strong magnetic field and the
anomaly of a $1+1$ dimensional massless Schwinger model on the one
hand and the nonplanar anomaly of a conventional noncommutative QED
on the other hand. Then in Sec. II, after giving a brief review on
the effective action of QED in a strong magnetic field, we will
derive the anomaly of QED in the LLL approximation and eventually
compare it with the nonplanar anomaly of the ordinary noncommutative
$U(1)$ gauge theory. In Sec. III, we will calculate the two-point
vertex function of the photon and determine the spectrum of a $3+1$
dimensional QED in the regime of LLL dominance. Sec. IV is devoted
to conclusions.

\subsection{Technical Details}
\subsection*{QED in a Strong Magnetic Field and Massless Schwinger
Model} \noindent The well established magnetic catalysis of
dynamical chiral symmetry breaking is a universal phenomenon in
$3+1$ dimensional QED in a strong constant magnetic field and leads
to a dimensional reduction $D\to D-2$ in a magnetic field. This is
why the $U_{A}(1)$ anomaly of $3+1$ dimensional QED is comparable
with the axial anomaly of a $1+1$ dimensional massless Schwinger
model
\begin{eqnarray}\label{AA-1}
\langle\partial_{\mu}j^{\mu}_{5}(x)\rangle=\frac{g}{2\pi}\epsilon_{\mu\nu}F^{\mu\nu}(x).
\end{eqnarray}
Recognizing this as the two dimensional version of the triangle
anomaly, the divergence of the axial vector current $j^{\mu}_{5}$ is
linear rather than quadratic in field strength tensor $F^{\mu\nu}$.
In the ordinary $1+1$ dimensional QED, it is easy to derive
(\ref{AA-1}) from the vacuum polarization tensor $\Pi_{\mu\nu}(q)$.
It is enough to use the relation between the vector current
$j^{\mu}$ and the axial vector current $j^{\mu}_{5}$ using the
properties of Dirac $\gamma$-matrices in two-dimensions,
$\gamma^{\mu}\gamma^{5}=-\epsilon^{\mu\nu}\gamma_{\nu}$, to get
\begin{eqnarray}\label{AB-1}
\langle j^{\mu}_{5}(q)\rangle&=&-\epsilon^{\mu\nu}\langle
j_{\nu}(q)\rangle\nonumber\\
&=&\epsilon^{\mu\nu}\
\frac{g}{\pi}\left(A_{\nu}(q)-\frac{q^{\nu}q^{\rho}}{q^{2}}A_{\rho}(q)\right).
\end{eqnarray}
Here, $j_{\nu}(q)$ is defined by the vacuum polarization tensor
$\Pi_{\mu\nu}(q)$
\begin{eqnarray}\label{AB-2}
\langle j_{\nu}(q)\rangle =\int dx\ e^{iqx}\langle
j_{\nu}(x)\rangle=-\frac{1}{g}\ \Pi_{\nu\rho}(q)\ A^{\rho}(q).
\end{eqnarray}
Using the methods familiar from four dimensions, the one-loop
contribution of $\Pi_{\mu\nu}(q)$ in two dimensions can be computed
and reads
\begin{eqnarray}\label{AB-3}
\Pi_{\mu\nu}(q)=\left(g_{\mu\nu}q^{2}-q_{\mu}q_{\nu}\right)\Pi(q^{2}),\qquad\qquad
\mbox{with}\qquad\qquad\Pi(q^{2})=\frac{g^{2}}{\pi}\
\frac{1}{q^{2}}.
\end{eqnarray}
On substituting back (\ref{AB-3}) into (\ref{AB-2}), the expression
on the second line of (\ref{AB-1}) is then found. Multiplying this
expression with $q_{\mu}$ yields the desired axial anomaly in two
dimensional momentum space
\begin{eqnarray}\label{AB-4}
q^{\mu}\langle j^{\mu}_{5}(q)\rangle= \frac{g}{\pi}
\epsilon^{\mu\nu} q_{\mu}A_{\nu}(q).
\end{eqnarray}
Transforming back into the coordinate space, the anomaly of an
ordinary massless Schwinger model is given by (\ref{AA-1}). In Sec.
II, we will use this method to determine the anomaly of QED in the
presence of a strong magnetic field in the momentum space. Assuming
that the constant magnetic field is directed in $x_{3}$ direction,
we find
\begin{eqnarray*}
q_{\mu}\langle
{\cal{J}}_{5}^{\mu}(q)\rangle=\frac{ieN_{f}|eB|\mbox{sign}(eB)}{2\pi^{2}}\
e^{-\frac{\mathbf{q}_{\perp}^{2}}{2|eB|}}\
\epsilon^{12ab}q_{a}A_{b}(\mathbf{q}_{\|},\mathbf{q}_{\perp}),\qquad
a,b=0,3,
\end{eqnarray*}
where the symbols $\perp$ and $\|$ are related to the $(1,2)$ and
$(0,3)$ components, respectively. To determine the anomaly in the
coordinate space, we will compactify two transverse coordinates
$\mathbf{x}_{\perp}$ around a circle with radius $R$ to study in
particular the role played by $\mathbf{q}_{\perp}=\mathbf{0}$.
Taking the decompactification limit $R\to \infty$, it turns out that
the zero transverse mode does not contribute to the unintegrated
form of the anomaly
\begin{eqnarray}\label{VV-7}
\langle\partial_{\mu}{\cal{J}}_{5}^{\mu}(x)\rangle=\frac{ieN_{f}|eB|\mbox{sign}(eB)}{2\pi^{2}}\
e^{\frac{\nabla_{\perp}^{2}}{2|eB|}}\
\epsilon^{12ab}\bar{F}_{ab}(\mathbf{x}_{\|},\mathbf{x}_{\perp}),
\end{eqnarray}
where
$\bar{F}_{ab}=\partial_{a}\bar{A}_{b}-\partial_{a}\bar{A}_{b}$, and
the nonzero transverse modes are defined by
$\bar{A}_{a}=A_{a}-A_{a}^{(0)}$. Here, the zero mode of the gauge
field $A_{a}^{(0)}$ is constant along the transverse directions
$\mathbf{x}_{\perp}$ and is defined by
$$
A_{a}^{(0)}(\mathbf{x}_{\|},\mathbf{q}_{\perp}=0)\equiv
\int\limits_{-R}^{+R}d^{2}x_{\perp}A^{(0)}_{a}(\mathbf{x}_{\|},\mathbf{x}_{\perp}).
$$
In Sec. III, we will calculate the 1PI effective action for two
photons in the LLL approximation, and determine the vacuum
polarization tensor of $3+1$ dimensional QED in a strong magnetic
field. In particular, we will show that the spectrum of this theory
consists of a massive photon of mass $M^{2}_{\gamma}\sim e^{2}|eB|$.
This is again in analogy to what happens in the ordinary Schwinger
model whose free neutral boson picks up a mass
$m_{\gamma}=\frac{g}{\sqrt{\pi}}$.\footnote{According to
\cite{schwinger}, the photon mass $m_{\gamma}$ in the Schwinger
model is one-loop exact.} Indeed, the emergence of a finite mass
arising from the 1PI effective action of two photons in LLL
approximation confirms the previous results from \cite{gusynin-1,
loskutov}. There, the photon mass of QED in a magnetic field was
calculated from the photon propagator ${\cal{D}}_{\mu\nu}$ of QED in
one-loop approximation with fermions from the LLL
\begin{eqnarray}
i{\cal{D}}_{\mu\nu}(q)=\frac{g_{\mu\nu}^{\perp}}{q^{2}}+\frac{q_{\mu}^{\|}q_{\nu}^{\|}}{q^{2}\mathbf{q}_{\|}^{2}}+
\frac{\left(g_{\mu\nu}^{\|}-q_{\mu}^{\|}q_{\nu}^{\|}/\mathbf{q}_{\|}^{2}\right)}{q^{2}+\mathbf{q}_{\|}^{2}
\Pi(\mathbf{q}_{\perp}^{2},\mathbf{q}_{\|}^{2})}-{\xi}\
\frac{q_{\mu}q_{\nu}}{(q^{2})^{2}},
\end{eqnarray}
with $\xi$ an arbitrary gauge parameter. Here,
$\Pi(\mathbf{q}_{\perp}^{2},\mathbf{q}_{\|}^{2})$ is given by
$\Pi(\mathbf{q}_{\perp}^{2},\mathbf{q}_{\|}^{2})=e^{-\frac{\mathbf{q}_{\perp}^{2}}{2|eB|}}\Pi(\mathbf{q}_{\|}^{2})$,
and $\Pi(\mathbf{q}_{\|}^{2})$ is calculated in \cite{gusynin-1,
loskutov} explicitly. As it turns out, since the LLL fermions couple
only to the longitudinal $(0,3)$ components of the photon fields, no
polarization effects are present in the transverse $(1,2)$ component
of ${\cal{D}}_{\mu\nu}(q)$. Therefore as the full propagator, one
can take the Feynman-like noncovariant propagator
\begin{eqnarray*}
{\cal{D}}_{\mu\nu}(q)=i\frac{g_{\mu\nu}^{\|}}{q^{2}+\mathbf{q}_{\|}^{2}\Pi\left(\mathbf{q}^{2}_{\|},\mathbf{q}_{\perp}^{2}\right)}.
\end{eqnarray*}
\par
\noindent It was shown in \cite{gusynin-1} that the kinematic region
mostly responsible for generating the fermion mass  is that with the
dynamical mass $m_{dyn}$ satisfying $m_{dyn}^{2}\ll
|\mathbf{q}_{\|}^{2}|\ll |eB|$ and $|\mathbf{q}_{\perp}^{2}|\ll
|eB|$. In that region, which is indeed the relevant regimes for the
LLL approximation, the fermions can be treated as massless
\cite{miransky-1} and the polarization operator can be calculated in
one-loop approximation. Here, one uses the asymptotic behavior of
$\Pi(\mathbf{q}_{\|}^{2})$ \cite{gusynin-1}, {\it i.e.},
\begin{eqnarray}
\Pi(\mathbf{q}_{\|}^{2})\simeq\  +\frac{N_{f}\alpha_{b}\ |eB|}{3\pi\
m_{dyn}^{2}}&\qquad\mbox{for}\qquad&m_{dyn}^{2}\gg
|\mathbf{q}_{\|}^{2}|,\label{AB-5a}\\
\Pi(\mathbf{q}_{\|}^{2})\simeq -\frac{2N_{f}\alpha_{b}\ |eB|}{\pi\
\mathbf{q}_{\|}^{2}}&\qquad\mbox{for}\qquad&m_{dyn}^{2}\ll
|\mathbf{q}_{\|}^{2}|,\label{AB-5b}
\end{eqnarray}
with $N_{f}$  the number of flavors and $\alpha_{b}\equiv
\frac{e_{b}^{2}}{4\pi}$ the running coupling.  Hence (\ref{AB-5b})
implies that
$$ \frac{1}{q^{2}+\mathbf{q}_{\|}^{2}\Pi(\mathbf{q}_{\|}^{2},\mathbf{q}_{\perp}^{2})}\simeq
\frac{1}{q^{2}-M_{\gamma}^{2}},\qquad\mbox{with}\qquad
M_{\gamma}^{2}=\frac{2N_{f}\alpha_{b} |eB|}{\pi}.
$$
The appearance of a finite photon mass is indeed a reminiscent of
the Higgs effect in $1+1$ dimensional Schwinger model
\cite{schwinger}. Note that although the IR dynamics of QED in the
presence of a magnetic field is very different from that in the
Schwinger model, the tensor and spinor structure of this dynamics is
exactly the same as in the Schwinger model \cite{gusynin-1}.
\subsection*{Anomalies of Noncommutative QED
and $U_{A}(1)$ Anomaly in a Strong Magnetic Field} \noindent The
main observation in this paper is related to the connection of the
$U_{A}(1)$ anomaly of QED in a strong magnetic field and the {\it
nonplanar} anomaly of the ordinary noncomutative $U(1)$ gauge
theory. At this stage, before describing the similarities between
these two anomalies, it will be instructive to summarize some of the
previous results of the anomalies in noncommutative QED \cite{ned-1,
ned-2, ned-3, ned-4}\footnote{Other aspects of the anomalies of NCFT
have been studied in \cite{nc-anomalies}. For a more complete list
of references see also \cite{ned-4}.}:
\par
As is well-known, noncommutative field theory (NCFT) is
characterized by a $\star$-deformation of the ordinary commutative
field theory. The noncommutative Moyal $\star$-product is defined by
\begin{eqnarray}\label{XX-5}
(f\star g)(x)\equiv f(x+\xi)\ \exp\left(\frac{i\Theta^{\mu\nu}}{2}\
\frac{\partial}{\partial\xi^{\mu}}\frac{\partial}{\partial\zeta^{\nu}}\right)
g(x+\zeta)\Bigg|_{\xi=\zeta=0}.
\end{eqnarray}
Due to the specific structure of the $\star$-product, noncommutative
field theories can be regarded as nonlocal field theories involving
higher order derivatives between the interacting fields.
Perturbatively, the theory consists therefore of planar and
nonplanar diagrams. The latter are usually the source of the
appearance of the above mentioned UV/IR mixing phenomenon, which is
shown to modify the anomalies in noncommutative field theory too
\cite{ned-2,ned-3}. The main observation in \cite{ned-1} and
\cite{ned-2} was that the noncommutative $U(1)$ gauge theory
consists of two different {\it global} axial vector currents; the
covariant $J^{\mu}_{5}=\psi_{\beta}\star\bar{\psi}_{\alpha}\
(\gamma^{\mu}\gamma^{5})^{\alpha\beta}$, and the invariant axial
vector current
$j^{\mu}_{5}=\bar{\psi}_{\alpha}\star\psi_{\beta}(\gamma^{\mu}\gamma^{5})^{\alpha\beta}.$
Naively, one would expect that these two currents have the same
anomaly. But, as it is shown in \cite{ned-4}, due to the properties
of $\star$-product, only the {\it integrated} form of two anomalies
are the same
\begin{eqnarray}\label{XX-2}
\int d^{2}x_{\perp} \langle D_{\mu}J^{\mu}_{5}(x)\rangle=\int
d^{2}x_{\perp} \langle\partial_{\mu}j^{\mu}_{5}(x)\rangle,
\end{eqnarray}
with $\mathbf{x}_{\perp}$ denoting the noncommutative directions.
Their {\it unintegrated} forms are indeed different; while the
anomaly corresponding to the covariant current, arising from the
planar diagrams of the theory, is the expected $\star$-deformation
of the ABJ anomaly \cite{ned-1}
\begin{eqnarray}\label{XX-6}
\langle D_{\mu}J_{5}^{\mu}(x)\rangle=-\frac{e^{2}}{16\pi^{2}}\
F_{\mu\nu}(x)\star\tilde{F}^{\mu\nu}(x),
\end{eqnarray}
the anomaly in $\langle\partial_{\mu}j^{\mu}_{5}\rangle$, receives
contribution from nonplanar diagrams and is therefore affected by
the noncommutative UV/IR mixing. The nonplanar (invariant) anomaly
of $\langle\partial_{\mu}j^{\mu}_{5}\rangle$ is calculated in
\cite{ned-2,ned-3} using various regularization methods.  Quoting in
particular the result from \cite{ned-3}, where the nonplanar anomaly
is calculated using the well-known Fujikawa's  path integral method,
the divergence of the invariant axial vector current is
\begin{eqnarray}\label{XX-1}
\langle\partial_{\mu}j^{\mu}_{5}(x)\rangle=\lim\limits_{M\to\infty}\
-\frac{e^{2}}{16\pi^{2}}\int\frac{d^{4}k}{(2\pi)^{4}}\frac{d^{4}p}{(2\pi)^{4}}\
e^{-\frac{M^{2}\tilde{q}^{2}}{4}}e^{-ikx}
F_{\mu\nu}(k)\frac{\sin(k\times p)}{k\times
p}\tilde{F}^{\mu\nu}(p)e^{-ipx}+\cdots,
\end{eqnarray}
where $M$ is the UV regulator. Here, we have used the notations
$q\equiv k+p$, $\tilde{q}^{\mu}\equiv \Theta^{\mu\nu}q_{\nu}$ and
$k\times p\equiv k_{\mu}\tilde{p}^{\mu}/2$. To show the celebrated
UV/IR mixing in the case of nonplanar anomaly (\ref{XX-1}), we have
considered two limits ${\tilde{q}^{2}}\gg \frac{1}{M^{2}}$ and
${\tilde{q}^{2}}\ll \frac{1}{M^{2}}$, separately. As it turns out
the limit ${\tilde{q}^{2}}\gg \frac{1}{M^{2}}$ is equivalent with
taking first the limit $M\to\infty$ and then $|\tilde{q}|\to 0$. In
this case, even before taking $|\tilde{q}|\to 0$, the anomaly
vanishes. In the opposite case, {\it i.e.} by taking first
$|\tilde{q}|\to 0$ and then $M\to \infty$, a finite anomaly arises.
Note that this limit can be understood as the limit
${\tilde{q}^{2}}\ll \frac{1}{M^{2}}$. In this case the exponent
$\exp\left(-\frac{M^{2}\tilde{q}^{2}}{4}\right)\to 1$ and we are
left with a finite nonplanar anomaly,
\begin{eqnarray}\label{A-1}
\langle\partial_{\mu}j^{\mu}_{5}\rangle=-\frac{e^{2}}{16\pi^{2}}\
F_{\mu\nu}\star' \tilde{F}^{\mu\nu}+\cdots,
\end{eqnarray}
where the generalized $\star'$-product is defined by
\begin{eqnarray}
(f\star' g)(x)\equiv
f(x+\xi)\frac{\sin\left(\frac{\Theta_{\mu\nu}}{2}\frac{\partial}{\partial\xi}\frac{\partial}{\partial\zeta}\right)}
{\frac{\Theta_{\mu\nu}}{2}\frac{\partial}{\partial\xi}\frac{\partial}{\partial\zeta}}g(x+\zeta)\bigg|_{\xi=\zeta=0}.
\end{eqnarray}
The ellipsis in (\ref{XX-1}) and (\ref{A-1}) indicate the
contribution of the expansion of a noncommutative Wilson line, which
is to be attached to $F_{\mu\nu}$ and $\widetilde{F}_{\mu\nu}$ in
order to restore the $\star$-gauge invariance of the result
\cite{ned-3}.
\par
In \cite{ned-4} we argue that the above results remain only correct
when we assume that the noncommutative $U(1)$ gauge theory is an
effective field theory which consists of a natural, large but finite
cutoff $M$. However, considering the noncommutative field theory as
a fundamental field theory and taking the limit of $M\to \infty$
first, the nonplanar anomaly vanishes except for the one point in
the momentum space, $|\tilde{q}|=0$; As it can easily checked in
(\ref{XX-1}), in this case, the phase factor
$\exp\left(-M^{2}\tilde{q}^{2}/{4}\right)=1$, and this leads to a
finite nonplanar anomaly. This is in accordance with the arguments
in \cite{armoni}, where it is emphasized that a nonvanishing
nonplanar anomaly is indeed necessary to guarantee that the
covariant and invariant currents have the same {\it integrated}
axial anomaly [see the argument leading to (\ref{XX-2})].
\par
To compute the nonplanar anomaly in such a fundamental theory,
without any natural cutoff, it is necessary to perform in addition
to the familiar UV regularization, an appropriate IR regularization.
In \cite{ned-4}, the IR regulator is introduced by compactifying
each space coordinates to a circle with radius $R$. Assuming that
the noncomutativity is between the spacial coordinates ${\mathbf
x}_{\perp}=(x_{1}, x_{2})$, and denoting the other two directions by
${\mathbf {x}}_{\|}=(x_{0},x_{3})$, the unintegrated form of the
nonplanar (invariant) anomaly is given by
\begin{eqnarray}\label{VV-1}
\langle\partial_{\mu}j^{\mu}_{5}(x)\rangle=-\frac{e^{2}}{16\pi^{2}}\
\frac{1}{(2R)^{2}}\int\limits_{-R}^{+R} d^{2}y_{\perp} \
F^{\alpha\beta}(\mathbf{x}_{\|},\mathbf{y}_{\perp})\tilde{F}_{\alpha\beta}(\mathbf{x}_{\|},\mathbf{y}_{\perp}).
\end{eqnarray}
Here, taking the decompactification (IR) limit $R\to\infty$ the
anomaly ``density'' vanishes due to $1/R^{2}$ dependence on the
r.h.s. of (\ref{VV-1}). To obtain the desired finite result, we
should integrate both sides over the noncommutative directions
$\mathbf{x}_{\perp}$ -- this removes the $R$ dependence on the
r.h.s. of (\ref{VV-1}) -- and then, take the limit $R\to\infty$.
This situation is as if a finite charge is evenly distributed over
the space giving zero density but still being totally nonzero
\cite{ned-4}. The integrated form of the nonplanar (invariant)
anomaly becomes
\begin{eqnarray}\label{XY-2}
\int\limits_{-R}^{+R}d^{2}x_{\perp}\
\langle\partial_{\mu}j^{\mu}_{5}(x)\rangle=-\frac{e^{2}}{16\pi^{2}}\
\frac{1}{(2R)^{2}}
\int\limits_{-R}^{+R}d^{2}x_{\perp}\int\limits_{-R}^{+R}
d^{2}y_{\perp} \
F^{\alpha\beta}(\mathbf{x}_{\|},\mathbf{y}_{\perp})\tilde{F}_{\alpha\beta}(\mathbf{x}_{\|},\mathbf{y}_{\perp}),
\end{eqnarray}
which survives the limit $R\to\infty$, {\it i.e.}
\begin{eqnarray}\label{XX-3}
\int\limits_{-\infty}^{+\infty}d^{2}x_{\perp}\
\langle\partial_{\mu}j^{\mu}_{5}(x)\rangle=-\frac{e^{2}}{16\pi^{2}}\
\int\limits_{-\infty}^{+\infty}d^{2}y_{\perp} \
F^{\alpha\beta}(\mathbf{x}_{\|},\mathbf{y}_{\perp})\tilde{F}_{\alpha\beta}(\mathbf{x}_{\|},\mathbf{y}_{\perp}).
\end{eqnarray}
Thus, the expression for the nonplanar anomaly turns out to be
independent of noncommutative coordinates $\mathbf{x}_{\perp}$.
Although this can be interpreted as a dimensional reduction in the
space-time coordinates, but the dimensional reduction seems to be
not complete here. This is because the nonplanar anomaly
(\ref{XX-3}) depends, as in any ordinary $3+1$ dimensional field
theory, quadratically on the field strength tensor. This is in
contrast to our result (\ref{VV-7}) on the anomaly of QED in the
presence of a strong magnetic field, which depends, as in a two
dimensional theory, linearly on the field strength tensor, at least
in the one-loop level. Further comparison shows that while the {\it
unintegrated} form of the nonplanar anomaly (\ref{VV-1}) receives
contribution {\it only} from zero noncommutative mode of the Fourier
transformed of ${\cal{F}}\equiv F\tilde{F}$,  the {\it unintegrated}
$U_{A}(1)$ anomaly of QED in the LLL approximation receives
additional contribution from nonzero transverse modes. To show this,
we have to compactify the transverse coordinates along a circle with
radius $R$. In the decompactification limit $R\to\infty$, the zero
transverse mode reappears in the {\it integrated} axial anomaly of
QED in a strong magnetic field, {\it i.e.} we have
\begin{eqnarray}
\int\limits_{-\infty}^{+\infty}d^{2}x_{\perp}\langle
\partial_{\mu}{\cal{J}}_{5}^{\mu}(x)\rangle=\frac{ieN_{f}|eB|\mbox{sign(eB)}}{4\pi^{2}}
\int\limits_{-\infty}^{+\infty}d^{2}x_{\perp}\epsilon^{12ab}F_{ab}(\mathbf{x}_{\|},\mathbf{x}_{\perp}),
\end{eqnarray}
with the field strength tensor $F_{ab}$ consisting of the nonzero
and zero transverse modes,  $F_{ab}=\bar{F}_{ab}+F_{ab}^{(0)}$. The
mechanism for the reappearance of the zero mode in the integrated
form of the QED anomaly in the LLL approximation is similar to what
happens in the noncommutative case [see how (\ref{XX-3}) arises from
(\ref{VV-1})].
\section{$U(1)$ axial anomaly in a strong magnetic field in the LLL approximation}
\setcounter{equation}{0}
\noindent In the first part of this section, we will briefly review
some results from \cite{miransky-1} on the effective action of QED
in a strong magnetic field. This will help us to set up our
notations. We then use the LLL fermion propagator to determine the
$U_{A}(1)$ anomaly of QED in a strong magnetic field.
\par
To put the QED dynamics in a magnetic field under control, we will
consider, as in \cite{miransky-1}, the case with a large number of
fermion flavors $N_{f}$. As is well-known the magnetic field is a
strong catalyst for dynamical chiral symmetry breaking and even the
weakest possible attraction between the fermions is enough for
dynamical mass generation. It is shown in \cite{gusynin-1} that the
dynamical mass behaves as $m_{dyn}\simeq
\sqrt{|eB|}\exp\left(-N_{f}\right)$ for a large running coupling
$\tilde{\alpha}_{b}\equiv N_{f}\alpha_{b}$. Thus, in the limit of
large $N_{f}$, the dynamical mass satisfies $m_{dyn}\ll
\sqrt{|eB|}$. This assumption guarantees in particular that no
dynamical symmetry breaking occurs, and as a consequence no light
(pseudo) Nambu-Goldstone bosons are produced. The low energy
effective theory will then consist only of photons and is given only
in terms of these fields. As for the current fermion mass $m$, it is
chosen to satisfy the condition $m\ll \sqrt{|eB|}$, which implies
that the magnetic field is very strong, and this is in fact a
guarantee that the LLL approximation is reliable.
\par
The effective action for photons is given by integrating out the
fermions and reads
\begin{eqnarray}\label{BB-1}
\Gamma=\Gamma^{(0)}+\Gamma^{(1)},\qquad
\Gamma^{(0)}=-\frac{1}{4}\int d^{4}x\ F_{\mu\nu}F^{\mu\nu},\qquad
\Gamma^{(1)}=-iN_{f}\ \mbox{Tr}\ln\left(iD\br-m\right),
\end{eqnarray}
with $D_{\mu}\equiv \partial_{\mu}-ieA_{\mu}$,
$F_{\mu\nu}=\partial_{\mu}A_{\nu}-\partial_{\nu}A_{\mu}$ and the
vector field $A_{\mu}=A_{\mu}^{cl.}+{\tilde{A}}_{\mu}$, where the
classical part $A_{\mu}^{cl.}\equiv \langle 0|A_{\mu}|0\rangle$ and
${\tilde{A}}_{\mu}$ is the fluctuating part. To proceed, it is
useful to choose the symmetric gauge
$$ A_{\mu}^{cl.}=\frac{B}{2}\left(0,x_{2},-x_{1},0\right). $$ This leads to
a magnetic field directed in $x_{3}$ dimensions. From now on, the
longitudinal $(0,3)$ directions are denoted by
$\mathbf{x}_{\|}=(x_{0},x_{3})$, and the transverse directions
$(1,2)$ by $\mathbf{x}_{\perp}=(x_{1},x_{2})$. Using the Schwinger
proper time formalism  \cite{schwinger-2}, it is possible to derive
the fermion propagator in this gauge. It is given by
\begin{eqnarray}\label{BB-2}
{\cal{S}}_{F}(x,y)&=&\exp\left(\frac{ie}{2}\left(x-y\right)^{\mu}A_{\mu}^{ext.}(x+y)\right)S(x-y)\nonumber\\
&=& e^{\frac{ieB}{2}\epsilon^{ab}x_{a}y_{b}} S(x-y),\qquad a,b=1,2,
\end{eqnarray}
where the first factor containing the external $A_{\mu}^{ext.}$ is
the Schwinger line integral \cite{schwinger-2}. Further, the Fourier
transform of the translationally invariant part $S(x-y)$ reads
\begin{eqnarray}\label{BB-3}
\tilde{S}(k)&=&i\int\limits_{0}^{\infty}ds\ e^{-ism^{2}}\
\exp\left(is\big[k_{\|}^{2}-\frac{k_{\perp}^{2}}{eBs\cot(eBs)}\big]\right)
\nonumber\\
&&\times \bigg\{\left(m+\mathbf{\gamma}^{\|}\cdot
\mathbf{k}_{\|}\right)\left(1-\gamma^{1}\gamma^{2}\tan(eBS)\right)-\mathbf{\gamma}^{\perp}\cdot
\mathbf{k}_{\perp}\left(1+\tan^{2}(eBs)\right)\bigg\},
\end{eqnarray}
where $\mathbf{k}_{\|}=(k_{0},k_{3})$ and
$\mathbf{\gamma}_{\|}=(\gamma_{0},\gamma_{3})$ and
$\mathbf{k}_{\perp}=(k_{1},k_{2})$ and
$\mathbf{\gamma}_{\perp}=(\gamma_{1},\gamma_{2})$.  After performing
the integral over $s$,  $\tilde{S}(k)$ can be decomposed as follows
\begin{eqnarray}\label{BB-4}
\tilde{S}(k)=ie^{-\frac{k_{\perp}^{2}}{|eB|}}\sum\limits_{n=0}^{\infty}(-1)^{n}\
\frac{D_{n}(eB,k)}{k_{\|}^{2}-m^{2}-2|eB|n},
\end{eqnarray}
with $D_{n}(eB,k)$ expressed through the generalized Laguerre
polynomials $L_{m}^{\alpha}$
\begin{eqnarray}\label{BB-5}
D_{n}(eB,k)=(\gamma^{\|}\cdot \mathbf{k}_{\|}+m)\ \bigg\{ 2\
{\cal{O}}\bigg[L_{n}\left(2\rho\right)-
L_{n-1}\left(2\rho\right)\bigg]+4\gamma^{\perp}\cdot
k_{\perp}L_{n-1}^{(1)}\left(2\rho\right) \bigg\}.
\end{eqnarray}
Here, we have introduced $\rho\equiv \frac{k_{\perp}^{2}}{|eB|}$ and
\begin{eqnarray}\label{BB-6}
{\cal{O}}\equiv\frac{1}{2}\left(1-i\gamma^{1}\gamma^{2}\mbox{sign}(eB)\right).
\end{eqnarray}
The lowest Landau level (LLL) is determined by $n=0$. Thus, the full
fermion propagator (\ref{BB-2}) in the LLL approximation can be
decomposed into two independent transverse and longitudinal parts
\cite{miransky-0}
\begin{nedalph}\label{BB-7a}
{\cal{S}}_{F}(x,y)=S_{\|}(\mathbf{x}_{\|}-\mathbf{y}_{\|})P(\mathbf{x}_{\perp},\mathbf{y}_{\perp})\
,
\end{eqnarray}
with the longitudinal part
\begin{eqnarray}\label{BB-7b}
S_{\|}(\mathbf{x}_{\|}-\mathbf{y}_{\|})=\int
\frac{d^{2}k_{\|}}{(2\pi)^{2}}\
e^{i\mathbf{k}_{\|}\cdot(\mathbf{x}-\mathbf{y})^{\|}}\
\frac{i{\cal{O}}}{\gamma^{\|}\cdot \mathbf{k}_{\|}-m},
\end{eqnarray}
and the transverse part
\begin{eqnarray}\label{BB-7c}
P(\mathbf{x}_{\perp},\mathbf{y}_{\perp})=\frac{|eB|}{2\pi}\exp\left(\frac{ieB}{2}\epsilon^{ab}x^{a}y^{b}-
\frac{|eB|}{4}\left(\mathbf{x}_{\perp}-\mathbf{y}_{\perp}\right)^{2}\right),\qquad
a,b=1,2.
\end{nedalph}

\par\noindent
Note that the longitudinal part (\ref{BB-7b}) is nothing else but
the fermion propagator in two dimensions. In particular, the matrix
${\cal{O}}$, defined in (\ref{BB-6}), is the projector on the
fermion (antifermion) states with the spin polarized along (opposite
to) the magnetic field \cite{miransky-0}. Further, the Schwinger
line integral included in the transverse part (\ref{BB-7c}) is
responsible for the noncommutative feature of the effective action
of QED in the LLL approximation
\begin{eqnarray}\label{BB-10}
\Gamma_{LLL}=\Gamma^{(0)}+\Gamma_{LLL}^{(1)},\qquad\Gamma_{LLL}^{(1)}=-\frac{iN_{f}\
|eB|}{2\pi}\int d^{2}x_{\perp}\ \mbox{Tr}_{\|}\left({\cal{O}}\
\ln\left(i\gamma^{\|}(\partial_{\|}-ie{\cal{A}}_{\|})\right)-m\right)_{\star},
\end{eqnarray}
where $\Gamma^{(0)}$ is defined in (\ref{BB-1}). Here, the $\star$
is the Moyal $\star$-product defined in (\ref{XX-5}). The appearance
of this product on the r.h.s. of the above equation shows that the
effective QED in the LLL dominant regime is indeed an effective
noncommutative field theory. In (\ref{BB-10}) the longitudinal {\it
smeared} gauge fields ${\cal{A}}_{\|}=({\cal{A}}_{0},{\cal{A}}_{3})$
is defined as
\begin{eqnarray}\label{BB-11}
{\cal{A}}_{\|}(x)\equiv
e^{\frac{\nabla_{\perp}^{2}}{4|eB|}}A_{\|}(x).
\end{eqnarray}
Note that here, since the one-loop contribution to the effective
action includes only the longitudinal ${\cal{A}}_{\|}$ field, the
spin structure of the LLL dynamics is $(1+1)$ dimensional
\cite{gusynin-1, miransky-1}, {\it i.e.} the LLL fermions couple
only to longitudinal components of the photon field. Note further
that the Gaussian-like form factor
$e^{{\nabla_{\perp}^{2}}/{4|eB|}}$ in the definition of the smeared
field arises essentially from the Schwinger line integral in the
transverse part of the fermion propagator (\ref{BB-7a}-c), and is
responsible for the noncommutative property of the effective action
in the regime of LLL dominance and the cancellation of the UV/IR
mixing \cite{minwalla} of the modified noncommutative field theory
\cite{miransky-1, miransky-2, miransky-0, ned-5}.
\par
Using these results, it is easy to calculate the $n$-point vertex
function of longitudinal photon in the LLL (from now on we will omit
the subscription $\|$ in the longitudinal gauge field
${\cal{A}}_{\|}$)
\begin{eqnarray*}
\Gamma_{LLL}^{(n)}=i\frac{(ie)^n\ N_{f}|eB|}{2\pi n}\int
d^{2}x_{\perp} d^{2}x_{1}^{\|}\cdots d^{2}x_{n}^{\|}\
\mbox{tr}\left(S_{\|}(\mathbf{x}_{1}^{\|}-\mathbf{x}_{2}^{\|})\widehat{{\cal{A}}}(\mathbf{x}_{}^{\perp},\mathbf{x}_{2}^{\|})\cdots
S_{\|}(\mathbf{x}_{n}^{\|}-\mathbf{x}_{1}^{\|})\widehat{{\cal{A}}}(\mathbf{x}_{}^{\perp},\mathbf{x}_{1}^{\|})
\right)_{\star},
\end{eqnarray*}
where $S_{\|}(\mathbf{x}_{\|}-\mathbf{y}_{\|})$ is defined in
(\ref{BB-7b}) and $\widehat{{\cal{A}}}\equiv
\gamma_{\|}\cdot{\cal{A}}^{\|}$. At this stage, we have all the
necessary tools to determine the ABJ anomaly in this modified
noncommutative $U(1)$ gauge theory.
\par
Let us now consider the axial vector current associated with the
$U_{A}(1)$ symmetry of the {\it original} QED Lagrangian
\begin{eqnarray}\label{BB-13}
{\cal{J}}_{5}^{\mu}(x)=\bar{\psi}(x)\gamma^{\mu}\gamma^{5}\psi(x).
\end{eqnarray}
To determine $\langle \partial_{\mu}{\cal{J}}^{\mu}_{5}(x)\rangle$
we will calculate first $\langle {\cal{J}}_{5}^{\mu}(q)\rangle$ in
the momentum space, and then, build $q_{\mu}\langle
{\cal{J}}_{5}^{\mu}\rangle$ in analogy to what is performed in
(\ref{AB-4}) to determine the anomaly in the ordinary two
dimensional Schwinger model from (\ref{AB-2}). To do so, we will use
the LLL fermion propagator (\ref{BB-7a}-c). Note that here, in
contrast to the ordinary $3+1$ dimensional QED gauge theory where a
triangle diagram of one axial and two vector current was responsible
for the emergence of the anomaly, the two-point function of
longitudinal photons which gives rise to the anomaly of our modified
noncommutative $U(1)$.  This is in analogy to the $1+1$ dimensional
QED and can be again regarded as a consequence of the dimensional
reduction in the presence of a strong magnetic background field.
\par
In the momentum space the vacuum expectation value of the axial
vector current ${\cal{J}}_{5}^{\mu}(x)$ is given by
\begin{eqnarray}\label{BB-14}
\langle {\cal{J}}_{5}^{\mu}(q)\rangle=\int d^{4}x\ e^{-iqx}\ \langle
{\cal{J}}_{5}^{\mu}(x)\rangle.
\end{eqnarray}
In the first order of perturbation theory, we have
\begin{eqnarray}\label{BB-15}
\langle {\cal{J}}_{5}^{\mu}(q)\rangle=-e\int d^{4}x\ d^{4}y\
e^{-iqx}\ \langle
\bar{\psi}(x)\gamma^{\mu}\gamma^{5}\psi(x)\bar{\psi}(y)\ {{A}}\br\
(y)\psi(y) \rangle,
\end{eqnarray}
that, after contacting the fermionic fields, leads to
\begin{eqnarray}\label{BB-16}
\langle {\cal{J}}_{5}^{\mu}(q)\rangle=-e\int d^{4}x\ d^{4}y\
e^{-iqx}\ \mbox{tr}\left( \gamma^{\mu}\gamma^{5}
S_{\|}(\mathbf{x}_{\|}-\mathbf{y}_{\|})P(\mathbf{x}_{\perp},\mathbf{y}_{\perp})\
{{A}}\br\ (y)\
S_{\|}(\mathbf{y}_{\|}-\mathbf{x}_{\|})P(\mathbf{y}_{\perp},\mathbf{x}_{\perp})\right).
\end{eqnarray}
Substituting $S_{\|}(\mathbf{x}_{\|}-\mathbf{y}_{\|})$ from
(\ref{BB-7b}), we get
\begin{eqnarray}\label{BB-17}
\langle {\cal{J}}_{5}^{\mu}(q)\rangle&=& -e\int d^{4}x\ d^{4}y\int
\frac{d^{4}p}{(2\pi)^{4}}{{A}}_{\nu}(p) e^{ipy}
\int\frac{d^{2}k_{\|}}{(2\pi)^{2}}\
\frac{d^{2}\ell_{\|}}{(2\pi)^{2}}\
e^{-iqx}e^{i\mathbf{k}_{\|}\cdot(\mathbf{x}-\mathbf{y})^{\|}}
e^{i\mathbf{\ell}_{\|}\cdot(\mathbf{y}-\mathbf{x})^{\|}}\nonumber\\
&&\times \mbox{tr}\left(\gamma^{\mu}\gamma^{5}
P(\mathbf{x}_{\perp},\mathbf{y}_{\perp})\ \frac{i}{\gamma^{\|}\cdot
\mathbf{k}_{\|}-m}{\cal{O}}\gamma^{\nu}P(\mathbf{y}_{\perp},\mathbf{x}_{\perp})\
\frac{i}{\gamma^{\|}\cdot \mathbf{\ell}_{\|}-m}{\cal{O}}\right).
\end{eqnarray}
Now using
\begin{eqnarray}\label{BB-18}
\int d^{2}x_{\perp} d^{2}y_{\perp}\ e^{-i\mathbf{q}_{\perp}\cdot
\mathbf{x}^{\perp}}\ P(\mathbf{x}_{\perp},\mathbf{y}_{\perp})
e^{i\mathbf{p}_{\perp}\cdot
\mathbf{y}^{\perp}}P(\mathbf{y}_{\perp},\mathbf{x}_{\perp})=2\pi|eB|\
\delta^{2}(\mathbf{p}_{\perp}-\mathbf{q}_{\perp})\
e^{-\frac{\mathbf{q}_{\perp}^{2}}{2|eB|}},
\end{eqnarray}
and performing the integration over $\mathbf{x}_{\|}$ and
$\mathbf{y}_{\|}$ coordinates, we arrive first at
\begin{eqnarray}\label{BB-19}
\langle {\cal{J}}_{5}^{\mu}(q)\rangle=\frac{e|eB|}{2\pi}\
e^{-\frac{\mathbf{q}_{\perp}^{2}}{2|eB|}} A_{\nu}(q)\int
\frac{d^{2}k_{\|}}{(2\pi)^{2}}\
\frac{\mbox{tr}\left(\gamma^{\mu}\gamma^{5}\left(\mathbf{k}\br\
_{\|}+m\right){\cal{O}}\gamma^{\nu}\left(\mathbf{k}\br\
_{\|}-\mathbf{q}\br\ _{\|}+m\right){\cal{O}}\right)}
{\left(\mathbf{k}_{\|}^{2}-m^{2}\right)\left(\left(\mathbf{k}_{\|}-\mathbf{q}_{\|}\right)^{2}-m^{2}\right)}.
\end{eqnarray}
To calculate this integral, let us first concentrate on the
expression in the nominator. Using the property of the matrix
${\cal{O}}$ as the projector in the longitudinal direction,
${\cal{O}}\gamma{\cal{O}}={\cal{O}}\gamma_{\|}$, we get
\begin{eqnarray*}
\lefteqn{\hspace{-2cm}\mbox{tr}\left(\gamma^{\mu}\gamma^{5}\left(\mathbf{k}\br\
_{\|}+m\right){\cal{O}}\gamma^{\nu}\left(\mathbf{k}\br\
_{\|}-\mathbf{q}\br\
_{\|}+m\right){\cal{O}}\right)A_{\nu}(q)=\mbox{tr}\left(\gamma^{\mu}
\gamma^{5}(\mathbf{k}\br\ _{\|}+m)\ \gamma_{\|}
^{\nu}(\mathbf{k}\br\ _{\|}-\mathbf{q}\br\ _{\|}+m)\
{\cal{O}}\right) A_{\nu}^{\|}(q)
}\nonumber\\
&&\hspace{2cm}=\frac{1}{2}\mbox{tr}\left(\gamma^{\mu}\gamma^{5}(\mathbf{k}\br\
_{\|}+m)\gamma_{\|}^{\nu}(\mathbf{k}\br _{\|}-\mathbf{q}\br\
_{\|}+m)\right)A_{\nu}^{\|}(q)\nonumber\\
&&\hspace{2.5cm}-\frac{i}{2}\ \mbox{sign}(eB)\
\mbox{tr}\left(\gamma^{\mu}\gamma^{5}(\mathbf{k}\br\
_{\|}+m)\gamma_{\|}^{\nu}(\mathbf{k}\br\ _{\|}-\mathbf{q}\br\
_{\|}+m)\gamma^{1}\gamma^{2}
 \right)A_{\nu}^{\|}(q).
\end{eqnarray*}
Here, we have used the definition of ${\cal{O}}$ in (\ref{BB-6}). To
calculate the traces of Dirac $\gamma$-matrices, we use the
relations
$\mbox{tr}\left(\gamma^{5}\gamma^{\alpha}\gamma^{\beta}\gamma^{\rho}\gamma^{\sigma}\right)=4i\epsilon^{\alpha\beta\rho\sigma}$,
and
\begin{eqnarray}
\mbox{tr}\left(\gamma^{5}\gamma^{\alpha}\gamma^{\beta}\gamma^{\rho}\gamma^{\sigma}\gamma^{\lambda}\gamma^{\tau}\right)=4ig_{\eta\xi}\left(
\epsilon^{\alpha\beta\rho\eta}s^{\sigma\lambda\tau\xi}-\epsilon^{\sigma\lambda\tau\eta}s^{\alpha\beta\rho\xi}
\right),
\end{eqnarray}
with $s^{\alpha\beta\rho\sigma}\equiv
g^{\alpha\beta}g^{\rho\sigma}-g^{\alpha\rho}g^{\beta\sigma}+g^{\alpha\sigma}g^{\beta\rho}$.
After some straightforward calculation, (\ref{BB-19}) can be written
as
\begin{eqnarray}\label{BB-20}
\lefteqn{\hspace{-1.5cm}\langle
{\cal{J}}_{5}^{\mu}(q)\rangle=-\frac{e|eB|\mbox{sign}(eB)}{\pi}\
e^{-\frac{\mathbf{q}_{\perp}^{2}}{2|eB|}} A_{b}(q)}\nonumber\\
&&\times \int\frac{d^{2}k_{\|}}{(2\pi)^{2}}\
\frac{\bigg\{\epsilon^{12\mu
b}\left(m^{2}-\mathbf{k}_{\|}\cdot(\mathbf{k}-\mathbf{q})^{\|}\right)+\left(\epsilon^{12\mu
c}g^{ab}+g^{bc}\epsilon^{12\mu
a}\right)\mathbf{k}_{a}^{\|}(\mathbf{k}-\mathbf{q})_{c}^{\|}\bigg\}}{(\mathbf{k}_{\|}^{2}-m^{2})\left((\mathbf{k}_{\|}
-\mathbf{q}_{\|})^{2}-m^{2}\right)},
\end{eqnarray}
where we have omitted the symbol $\|$ on the gauge field $A_{\mu}$
and the metric $g_{\mu\nu}$. Instead, we have used the indices
$a,b,c=0,3,$ to denote the projection into the longitudinal
directions $\mathbf{x}_{\|}=(x_{0},x_{3})$. Note that due to the
antisymmetry of the $\epsilon^{\alpha\beta\gamma\rho}$ tensor, $\mu$
in $\epsilon^{12\mu c}$ on the r.h.s. must be chosen to be
$\mu=0,3$. This is indeed the first signature for the dimensional
reduction from $D=4$ to $D=2$ dimensions in the result of the
anomaly. The rest of the calculation is a straightforward
computation of the two dimensional Feynman integral over
$\mathbf{k}_{\|}$. Introducing first the Feynman parameter $0\leq
\alpha\leq 1$ and then performing a finite shift of integration
$\mathbf{k}_{\|}\to \mathbf{k}_{\|}+\alpha \mathbf{q}_{\|}$, we
arrive first at
\begin{eqnarray}\label{BB-21}
\langle
{\cal{J}}_{5}^{\mu}(q)\rangle&=&-\frac{e|eB|\mbox{sign}(eB)}{\pi}\
e^{-\frac{\mathbf{q}_{\perp}^{2}}{2|eB|}}\epsilon^{12\mu
a}A^{b}(q)\int\limits_{0}^{1}
d\alpha\bigg[\int\frac{d^{2}k_{\|}}{(2\pi)^{2}}\
\frac{2\mathbf{k}_{a}^{\|}\mathbf{k}_{b}^{\|}}{(\mathbf{k}^{2}_{\|}-\Delta)^{2}}-\int\frac{d^{2}k_{\|}}{(2\pi)^{2}}\
\frac{g_{ab}}{(\mathbf{k}_{\|}^{2}-\Delta)}\bigg]\nonumber\\
&&-\frac{2e|eB|\mbox{sign}(eB)}{\pi}\
e^{-\frac{\mathbf{q}_{\perp}^{2}}{2|eB|}}\epsilon^{12\mu
a}A^{b}(q)(\mathbf{q}_{\|}^{2}g_{ab}-\mathbf{q}_{a}^{\|}\mathbf{q}_{b}^{\|})\
\int\limits_{0}^{1}d\alpha\
\alpha(1-\alpha)\int\frac{d^{2}k_{\|}}{(2\pi)^{2}}\
\frac{1}{(\mathbf{k}_{\|}^{2}-\Delta)^{2}},\nonumber\\
\end{eqnarray}
where $\Delta\equiv m^{2}-\alpha(1-\alpha)\mathbf{q}_{\|}^{2}$. As
for the first two integrals, it can be shown that although they are
both infinite in the UV limit $| \mathbf{k}_{\|}|\to\infty$, they
cancel each other, and we are left with
\begin{eqnarray}\label{BB-22}
\langle
{\cal{J}}_{5}^{\mu}(q)\rangle=-\frac{ie|eB|\mbox{sign}(eB)}{2\pi^{2}}\
e^{-\frac{\mathbf{q}_{\perp}^{2}}{2|eB|}}\epsilon^{12\mu
a}A^{b}(q)(\mathbf{q}_{\|}^{2}g_{ab}-\mathbf{q}_{a}^{\|}\mathbf{q}_{b}^{\|})\
\int\limits_{0}^{1}d\alpha\ \alpha(1-\alpha)\ \frac{1}{\Delta}.
\end{eqnarray}
In the chiral limit, $m\to 0$, using the definition of $\Delta$, we
get\footnote{The chiral limit is taken just to isolate the anomaly
in the divergence of the axial vector current arising from the
quantum effects. According to the arguments in \cite{miransky-1},
the fermions can be treated as massless in the regime of LLL
dominance.}
\begin{eqnarray}\label{BB-23}
\langle
{\cal{J}}_{5}^{\mu}(q)\rangle=+\frac{ie|eB|\mbox{sign}(eB)}{2\pi^{2}}\
e^{-\frac{\mathbf{q}_{\perp}^{2}}{2|eB|}}\epsilon^{12\mu
a}A^{b}(q)\left(g_{ab}-\frac{\mathbf{q}_{a}^{\|}\mathbf{q}_{b}^{\|}}{\mathbf{q}_{\|}^{2}}\right).
\end{eqnarray}
The ABJ anomaly of QED with $N_{f}$ flavors in a strong magnetic
field in the LLL approximation is then found by multiplying
(\ref{BB-23}) with $q_{\mu}$ and using the antisymmetry property of
$\epsilon^{12ab}$ tensor
\begin{eqnarray}\label{BB-24}
q_{\mu}\langle
{\cal{J}}_{5}^{\mu}(q)\rangle=+\frac{ieN_{f}|eB|\mbox{sign}(eB)}{2\pi^{2}}\
e^{-\frac{\mathbf{q}_{\perp}^{2}}{2|eB|}}\epsilon^{12a
b}q_{a}^{\|}A_{b}(\mathbf{q}_{\|},\mathbf{q}_{\perp}).
\end{eqnarray}
Before transforming this result back into the coordinate space, let
us compare it with (\ref{XX-1}), the nonplanar anomaly of the
invariant current $j_{5}^{\mu}\equiv
\psi^{\beta}\star\bar{\psi}^{\alpha}(\gamma^{\mu}\gamma^{5})_{\alpha\beta}$
of the ordinary noncommutative QED. Assuming that the
noncommutativity is defined only between two coordinates $x_{1}$ and
$x_{2}$, (\ref{XX-1}) can be written as
\begin{eqnarray}\label{BB-25}
q_{\mu}\langle j^{\mu}_{5}(q)\rangle=\lim\limits_{M\to\infty}\
-\frac{ie^{2}}{16\pi^{2}}\
e^{-\frac{(M\theta)^{2}\mathbf{q}_{\perp}^{2}}{4}}\int\frac{d^{4}p}{(2\pi)^{4}}\
F_{\mu\nu}(q-p)\frac{\sin(q\times p)}{q\times
p}\tilde{F}^{\mu\nu}(p)+\cdots,
\end{eqnarray}
with $q\equiv k+p$. Here, we have transformed (\ref{XX-1}) into the
momentum space and replaced the phase factor
$e^{-M^{2}\tilde{q}^{2}/4}$ with
$\tilde{q}_{\mu}\equiv\Theta_{\mu\nu}q^{\nu}$ by
$e^{-(M\theta)^{2}\mathbf{q}_{\perp}^{2}/4}$, where $\theta$ is
defined by $\Theta_{ij}\equiv\theta \epsilon_{ij}$, with $i,j=1,2$
-- this gives us the possibility to compare (\ref{BB-25}) with
(\ref{BB-24}). As we have argued in Sec. I, the phase factor
$e^{-(M\theta)^{2}\mathbf{q}_{\perp}^{2}/4}$ is indeed the origin of
the appearance of UV/IR mixing; Assuming that the ordinary
noncommutative $U(1)$ gauge theory is a fundamental theory and
taking the limit $M\to\infty$, the nonplanar anomaly vanishes
everywhere in the momentum space except for the point
$\mathbf{q}_{\perp}=\mathbf{0}$. In this case the phase factor
$$e^{-(M\theta)^{2}\mathbf{q}_{\perp}^{2}/4}=1,\qquad\mbox{for}\qquad
\mathbf{q}_{\perp}=\mathbf{0},$$ and the nonplanar anomaly turns out
to be given by (\ref{VV-1}), where a compactification around a
circle with the radius $R$ is performed. Only in this way it could
be shown in \cite{ned-4} that although the {\it unintegrated} form
of the nonplanar anomaly vanishes in the decompactification limit
$R\to\infty$, the {\it integrated} form of the nonplanar anomaly is
finite and is given by (\ref{XX-3}). Note that (\ref{VV-1}) can also
be interpreted as if the unintegrated nonplanar anomaly receives a
finite contribution {\it only} from the zero mode of the Fourier
transformed of ${\cal{F}}\equiv F_{\mu\nu}\tilde{F}^{\mu\nu}$ in the
noncommutative coordinates $\mathbf{x}_{\perp}$, {\it i.e.} from
\begin{eqnarray}\label{DD-1}
\langle \partial_{\mu}j_{5}^{\mu} (x)\rangle \sim
\widetilde{\cal{F}}(\mathbf{x}_{\|},
\mathbf{q}_{\perp}=\mathbf{0})\equiv
\frac{1}{(2R)^{2}}\int\limits_{-R}^{+R}d^{2}x_{\perp}\
{\cal{F}}(\mathbf{x}_{\|},\mathbf{x}_{\perp}).
\end{eqnarray}
This contribution vanishes in the $R\to\infty$ limit.
\par
The above analysis of the nonplanar anomaly of the ordinary
noncommutative QED shows the special role played by
$\mathbf{q}_{\perp}=\mathbf{0}$ in the expression (\ref{BB-24}) for
the anomaly of QED in a strong magnetic field in the LLL
approximation. Note that in this case the magnetic field provides a
natural cutoff for the modified noncommutative QED in the LLL
approximation, and can be compared with the product of the UV cutoff
$M$ and the IR cutoff $\theta$ of the ordinary noncommutative field
theory; $\sqrt{|eB|}\sim (M\theta)^{-1}$, where here, in contrast to
the ordinary noncommutative case, {\it both} $M$ and $\theta$ are
kept finite, so that $\mathbf{q}_{\perp}^{2}\ll |eB|$ is correct and
thus the reliability of the LLL approximation is guaranteed.
\par
To transform (\ref{BB-24}) back into the coordinate space and to be
specially careful about the role played by
$\mathbf{q}_{\perp}=\mathbf{0}$, we compactify, as in the case of
the ordinary noncommutativity, two transverse coordinates
$\mathbf{x}_{\perp}$ around a circle with radius $R$. Multiplying
both sides of (\ref{BB-24}) with $e^{iqx}$ and integrating (summing)
over the continuous (discrete) momenta $q_{\|}$ ($q_{\perp}$), we
have first
\begin{eqnarray}\label{BB-26}
\langle
\partial_{\mu}{\cal{J}}_{5}^{\mu}(x)\rangle=-\frac{eN_{f}|eB|\mbox{sign}(eB)}{2\pi^{2}}\
\int\frac{d^{2}q_{\|}}{(2\pi)^{2}}\
\frac{1}{(2R)^{2}}\sum\limits_{\mathbf{q}_{\perp}}
e^{i(\mathbf{q}_{\|}\cdot
\mathbf{x}^{\|}+\mathbf{q}_{\perp}\cdot\mathbf{x}^{\perp})}\
e^{-\frac{\mathbf{q}_{\perp}^{2}}{2|eB|}}\epsilon^{12a
b}q_{a}A_{b}(\mathbf{q}_{\|},\mathbf{q}_{\perp}),
\end{eqnarray}
where we have used the relation
\begin{eqnarray}\label{YY-1}
\int\frac{d^{2}{q}_{\perp}}{(2\pi)^{2}}\to
\frac{1}{(2R)^{2}}\sum\limits_{\mathbf{q}_{\perp}}\
,\qquad\mbox{with} \qquad \mathbf{q}_{\perp}\equiv \frac{\pi
\mathbf{n_{q}}}{R}.
\end{eqnarray}
To proceed, we separate the sum over discrete transverse momenta
$\mathbf{q}_{\perp}$ into the nonzero $\mathbf{q}_{\perp}\neq
\mathbf{0}$ and the zero mode $\mathbf{q}_{\perp}=\mathbf{0}$, so
that (\ref{BB-26}) can be written as
\begin{eqnarray}\label{BB-27}
\langle
\partial_{\mu}{\cal{J}}_{5}^{\mu}(x)\rangle&=&-\frac{eN_{f}|eB|\mbox{sign}(eB)}{2\pi^{2}}\epsilon^{12a
b}\left( \int\frac{d^{2}q_{\|}}{(2\pi)^{2}}\
\frac{1}{(2R)^{2}}\sum\limits_{\mathbf{q}_{\perp}\neq \mathbf{0}}
e^{i(\mathbf{q}_{\|}\cdot\mathbf{x}^{\|}+\mathbf{q}_{\perp}\cdot\mathbf{x}^{\perp})}\
e^{-\frac{\mathbf{q}_{\perp}^{2}}{2|eB|}}q_{a}\bar{A}_{b}(\mathbf{q}_{\|},\mathbf{q}_{\perp}\neq
\mathbf{0})\right.\nonumber\\
&&\left. + \int\frac{d^{2}q_{\|}}{(2\pi)^{2}}\ \frac{1}{(2R)^{2}}
e^{i\mathbf{q}_{\|}\cdot\mathbf{x}^{\|}}
q_{a}{A}_{b}^{(0)}(\mathbf{q}_{\|},
\mathbf{q}_{\perp}=\mathbf{0})\right).
\end{eqnarray}
Substituting the Fourier transformed of the nonzero transverse modes
$\bar{A}$,
\begin{eqnarray}\label{BB-28}
\bar{A}_{b}(\mathbf{q}_{\|},\mathbf{q}_{\perp}\neq
\mathbf{0})=\int\limits_{-\infty}^{+\infty}
d^{2}y_{\|}\int\limits_{-R}^{+R} d^{2}y_{\perp}
\bar{A}_{b}(\mathbf{y}_{\|}, \mathbf{y}_{\perp})\
e^{-i(\mathbf{q}_{\|}\cdot \mathbf{y}^{\|}+\mathbf{q}_{\perp}\cdot
\mathbf{y}^{\perp})},
\end{eqnarray}
and the zero transverse modes ${A}^{(0)}$,
\begin{eqnarray}\label{BB-29}
{A}^{(0)}_{b}(\mathbf{q}_{\|},\mathbf{q}_{\perp}=\mathbf{0})=\int\limits_{-\infty}^{+\infty}
d^{2}y_{\|}\int\limits_{-R}^{+R} d^{2}y_{\perp}
{A}^{(0)}_{b}(\mathbf{y}_{\|}, \mathbf{y}_{\perp})\
e^{-i\mathbf{q}_{\|}\cdot \mathbf{y}^{\|}},
\end{eqnarray}
into the r.h.s. of (\ref{BB-27}) and using the identities
\begin{eqnarray*}
\int\frac{d^{2}q_{\|}}{(2\pi)^{2}}e^{i\mathbf{q}_{\|}\cdot(\mathbf{x}-\mathbf{y})^{\|}}=\delta^{2}(\mathbf{x}_{\|}-\mathbf{y}_{\|}),
\qquad\mbox{and}\qquad
\frac{1}{(2R)^{2}}\sum\limits_{\mathbf{q}_{\perp}\neq
0}e^{i\mathbf{q}_{\perp}\cdot(\mathbf{x}-\mathbf{y})^{\perp}}=\delta^{2}(\mathbf{x}_{\perp}-\mathbf{y}_{\perp}),
\end{eqnarray*}
we arrive after integrating over $\mathbf{y}_{\|}$ and
$\mathbf{y}_{\perp}$ at
\begin{eqnarray*}
\langle
\partial_{\mu}{\cal{J}}_{5}^{\mu}(x)\rangle&=&\frac{ieN_{f}|eB|\mbox{sign}(eB)}{2\pi^{2}}\left(
e^{\frac{\nabla_{\perp}^{2}}{2|eB|}}\epsilon^{12a
b}\partial_{a}\bar{A}_{b}(\mathbf{x}_{\|},\mathbf{x}_{\perp})+
\frac{1}{(2R)^{2}}\int\limits_{-R}^{+R} d^{2}y_{\perp} \epsilon^{12a
b}\partial_{a}{A}^{(0)}_{b}(\mathbf{x}_{\|},\mathbf{y}_{\perp})
\right).
\end{eqnarray*}
Using now the notations
$$ \bar{F}_{ab}\equiv \partial_{a}\bar{A}_{b}-\partial_{b}\bar{A}_{a}, \qquad \mbox{and} \qquad {F}^{(0)}_{ab}\equiv \partial_{a}{A}^{(0)}_{b}
-\partial_{b}A_{a}^{(0)},$$ the $U_{A}(1)$ anomaly of QED in a
strong magnetic field for finite compactification length $R$  reads
\begin{eqnarray}\label{BB-31}
\langle
\partial_{\mu}{\cal{J}}_{5}^{\mu}(x)\rangle=\frac{ieN_{f}|eB|\mbox{sign}(eB)}{4\pi^{2}}\left(
e^{\frac{\nabla_{\perp}^{2}}{2|eB|}}\epsilon^{12a
b}\bar{F}_{ab}(\mathbf{x}_{\|},\mathbf{x}_{\perp})+
\frac{1}{(2R)^{2}}\int\limits_{-R}^{+R} d^{2}y_{\perp} \epsilon^{12a
b}{F}^{(0)}_{ab}(\mathbf{x}_{\|},\mathbf{y}_{\perp}) \right).
\end{eqnarray}
Apart from the fact that the anomaly of $3+1$ dimensional QED in the
strong magnetic field is, in contrast to the ordinary noncommutative
QED, linear in $F_{\mu\nu}$, at least at this one loop level, the
above situation is the same as in (\ref{VV-1}), {\it i.e.} by taking
the decompactification limit $R\to\infty$, the contribution from the
zero mode vanishes and we are left with the $R$ independent first
term in (\ref{BB-31}) from the contribution of the nonzero
transverse modes to the anomaly
\begin{eqnarray}\label{BB-32}
\langle
\partial_{\mu}{\cal{J}}_{5}^{\mu}(x)\rangle=\frac{ieN_{f}|eB|\mbox{sign}(eB)}{4\pi^{2}}
e^{\frac{\nabla_{\perp}^{2}}{2|eB|}}\epsilon^{12a
b}\bar{F}_{ab}(\mathbf{x}_{\|},\mathbf{x}_{\perp}).
\end{eqnarray}
A finite nonvanishing contribution of the zero transverse mode to
the axial anomaly of QED in a strong magnetic field arises only when
we integrate, as in the case of nonplanar anomaly [see
(\ref{XY-2})-(\ref{XX-3})], over $\mathbf{x}_{\perp}$ on both sides
of (\ref{BB-31}).\footnote{In the case of nonplanar anomaly,
$\mathbf{x}_{\perp}=(x_{1},x_{2})$ are the noncommutative
directions.} In this way, the $R$-dependence in the second term of
(\ref{BB-31}) cancels and the integrated form of the anomaly of QED
in a strong magnetic field becomes
\begin{eqnarray*}
\lefteqn{\int\limits_{-R}^{+R}d^{2}x_{\perp}\langle
\partial_{\mu}{\cal{J}}_{5}^{\mu}(x)\rangle=}\nonumber\\
&=&\frac{ieN_{f}|eB|\mbox{sign}(eB)}{4\pi^{2}}\left(\int\limits_{-R}^{+R}d^{2}x_{\perp}
e^{\frac{\nabla_{\perp}^{2}}{2|eB|}}\epsilon^{12a
b}\bar{F}_{ab}(\mathbf{x}_{\|},\mathbf{x}_{\perp})+
\frac{1}{(2R)^{2}}\int\limits_{-R}^{+R}d^{2}x_{\perp}\int\limits_{-R}^{+R}
d^{2}y_{\perp} \epsilon^{12a
b}{F}^{(0)}_{ab}(\mathbf{x}_{\|},\mathbf{y}_{\perp}) \right),
\end{eqnarray*}
which survives even in  $R\to\infty$ limit, {\it i.e.},
\begin{eqnarray*}
\int\limits_{-\infty}^{+\infty}d^{2}x_{\perp}\langle
\partial_{\mu}{\cal{J}}_{5}^{\mu}(x)\rangle=\frac{ieN_{f}|eB|\mbox{sign}(eB)}{4\pi^{2}}\left(\int\limits_{-\infty}^{+\infty}d^{2}x_{\perp}
e^{\frac{\nabla_{\perp}^{2}}{2|eB|}}\epsilon^{12a
b}\bar{F}_{ab}(\mathbf{x}_{\|},\mathbf{x}_{\perp})+
\int\limits_{-\infty}^{+\infty} d^{2}y_{\perp} \epsilon^{12a
b}{F}^{(0)}_{ab}(\mathbf{x}_{\|},\mathbf{y}_{\perp}) \right).
\end{eqnarray*}
Expanding now the phase factor $e^{\nabla^{2}_{\perp}/2|eB|}$ and
neglecting the surface term arising from the term including the
transverse derivatives $\nabla_{\perp}$, we get
\begin{eqnarray}
\int\limits_{-\infty}^{+\infty}d^{2}x_{\perp}\langle
\partial_{\mu}{\cal{J}}_{5}^{\mu}(x)\rangle=\frac{ieN_{f}|eB|\mbox{sign}(eB)}{4\pi^{2}}\int\limits_{-\infty}^{+\infty}d^{2}x_{\perp}
\epsilon^{12a b}{F}_{ab}(\mathbf{x}_{\|},\mathbf{x}_{\perp}),
\end{eqnarray}
with $F_{ab}=\bar{F}_{ab}+F^{(0)}_{ab}$ including the nonzero
$\bar{F}_{ab}$ {\it and} the zero transverse modes $F_{ab}^{(0)}$.
Hence, in contrast to the nonplanar anomaly of the ordinary
noncommutative QED, where the integrated anomaly in the limit
$R\to\infty$ receives contribution {\it only} from the zero modes of
the fields in  noncommutative directions, here, the integrated form
of the axial anomaly of QED in the LLL approximation includes both
nonzero and zero transverse modes.
\section{Two-point vertex function of photons in the LLL approximation and the photon mass}
\setcounter{equation}{0} \noindent In this section, the two-point
effective vertex function of photons in the LLL approximation will
be determined. In  particular, the special role played by the zero
transverse mode of the photon field will be studied in detail. We
start with the expression of the two-point vertex function at
one-loop level
\begin{eqnarray}\label{CC-1}
\Gamma_{LLL}^{(2)}=-(ie)^{2}N_{f}\int d^{4}x\ d^{4}y \
\mbox{tr}\left({\cal{S}}_{F}(x,y)A\br\ (y){\cal{S}}_{F}(y,x)A\br\
(y-x)\right).
\end{eqnarray}
Substituting ${\cal{S}}_{F}(x,y)$ from (\ref{BB-7a}-c), using the
relation (\ref{BB-18}), and integrating over $\mathbf{x}_{\|}$ and
$\mathbf{y}_{\|}$, we arrive after a straightforward calculation at
\begin{eqnarray*}
\Gamma_{LLL}^{(2)}=-\frac{e^{2}N_{f}|eB|}{2\pi}\int
\frac{d^{4}q}{(2\pi)^{4}}\frac{d^{2}k_{\|}}{(2\pi)^{2}}\
e^{-\frac{\mathbf{q}_{\perp}^{2}}{2|eB|}}\
\mbox{tr}\left(\frac{(\mathbf{k}\br\
_{\|}+m)}{\mathbf{k}_{\|}^{2}-m^{2}}{\cal{O}}\gamma^{a}\frac{(\mathbf{k}\br\
_{\|}+\mathbf{q}\br\
_{\|}+m)}{\left((\mathbf{k}_{\|}+\mathbf{q}_{\|})^{2}-m^{2}\right)}\gamma^{\nu}\right)\
A_{a}(q)A_{\nu}(-q),
\end{eqnarray*}
with $a=0,3$ and $\nu=0,1,2,3$. Here, as in the previous section, we
have used the property of the ${\cal{O}}$ matrix (\ref{BB-6}),
${\cal{O}}\gamma^{\mu}{\cal{O}}={\cal{O}}\gamma^{m}$, with
$\mu=0,\cdots,3$ and $m=0,3$. Using further the definition of
${\cal{O}}$, and following the standard procedure to evaluate the
two-dimensional Feynman integrals, {\it i.e.} introducing the
Feynman parameter $\alpha$ and performing a shift of integration
variable $\mathbf{k}_{\|}\to \mathbf{k}_{\|}-\alpha\mathbf{q}_{\|}$,
we arrive first at
\begin{eqnarray}\label{CC-3}
\Gamma_{LLL}^{(2)}&=&-\frac{e^{2}N_{f}|eB|}{\pi}\int
\frac{d^{4}q}{(2\pi)^{4}}e^{-\frac{\mathbf{q}_{\perp}^{2}}{2|eB|}}\
A^{a}(q)A^{b}(-q)\int\limits_{0}^{1}
d\alpha\bigg[\int\frac{d^{2}k_{\|}}{(2\pi)^{2}}\
\frac{2\mathbf{k}_{a}^{\|}\mathbf{k}_{b}^{\|}}{(\mathbf{k}^{2}_{\|}-\Delta)^{2}}-\int\frac{d^{2}k_{\|}}{(2\pi)^{2}}\
\frac{g_{ab}}{(\mathbf{k}_{\|}^{2}-\Delta)}\bigg]\nonumber\\
&&-\frac{2e^{2}N_{f}|eB|}{\pi}\int \frac{d^{4}q}{(2\pi)^{4}}\
e^{-\frac{\mathbf{q}_{\perp}^{2}}{2|eB|}}A^{a}(q)A^{b}(-q)(\mathbf{q}_{\|}^{2}g_{ab}-\mathbf{q}_{a}^{\|}\mathbf{q}_{b}^{\|})\
\int\limits_{0}^{1}d\alpha\
\alpha(1-\alpha)\int\frac{d^{2}k_{\|}}{(2\pi)^{2}}\
\frac{1}{(\mathbf{k}_{\|}^{2}-\Delta)^{2}},\nonumber\\
\end{eqnarray}
where $\Delta\equiv m^{2}-\alpha(1-\alpha)\mathbf{q}_{\|}^{2}$.
Here, as in the ordinary $1+1$ dimensional Schwinger model, the
two-point vertex function (\ref{CC-3}) is closely related to
$\langle j_{5}^{\mu}(q)\rangle$ from (\ref{BB-21}). As in that case,
the first two integrals cancel each other and we are left with
\begin{eqnarray}\label{CC-4}
\Gamma_{LLL}^{(2)}=-\frac{ie^{2}N_{f}|eB|}{2\pi^{2}}\int
\frac{d^{4}q}{(2\pi)^{4}}\
e^{-\frac{\mathbf{q}_{\perp}^{2}}{2|eB|}}A^{a}(q)A^{b}(-q)(\mathbf{q}_{\|}^{2}g_{ab}-\mathbf{q}_{a}^{\|}\mathbf{q}_{b}^{\|})\
\int\limits_{0}^{1}d\alpha\ \alpha(1-\alpha)\ \frac{1}{\Delta}.
\end{eqnarray}
Taking again the chiral limit $m\to 0$, we get
\begin{eqnarray}\label{CC-5}
\Gamma_{LLL}^{(2)}=-\frac{ie^{2}N_{f}|eB|}{2\pi^{2}}\int
\frac{d^{4}q}{(2\pi)^{4}}\
e^{-\frac{\mathbf{q}_{\perp}^{2}}{2|eB|}}A^{a}(q)\left(g_{ab}-\frac{\mathbf{q}_{a}^{\|}\mathbf{q}_{b}^{\|}}{\mathbf{q}_{\|}^{2}}\right)A^{b}(-q).
\end{eqnarray}
To extract the one-loop vacuum polarization tensor $\Pi_{\mu\nu}$ of
the modified noncommutative QED from this result, we use, as was
suggested in \cite{miransky-1} and \cite{miransky-0}, the definition
of the smeared fields in the momentum space
\begin{eqnarray}\label{CC-6}
{\cal{A}}_{\mu}(q)\equiv e^{-\frac{\mathbf{q}^{2}_{\perp}}{4|eB|}}
A_{\mu}(q).
\end{eqnarray}
Using now the general relation between the two-point vertex function
and the vacuum polarization tensor of the smeared photon fields
\begin{eqnarray}\label{CC-7}
\Gamma^{(2)}_{LLL}=\int\frac{d^{4}q}{(2\pi)^{4}}\ {\cal{A}}^{a}(q)\
\left(-i\Pi_{ab}(\mathbf{q}_{\|})\right)\ {\cal{A}}^{b}(-q),
\end{eqnarray}
and comparing with (\ref{CC-5}), the projection of the vacuum
polarization tensor $\Pi_{\mu\nu}$ into the longitudinal directions,
$\Pi_{ab}$ with $a,b=0,3$, reads
\begin{eqnarray}\label{CC-8}
\Pi_{ab}(\mathbf{\mathbf{q}_{\|}})=\left(g_{ab}\mathbf{q}_{\|}^{2}-{\mathbf{q}_{a}^{\|}\mathbf{q}_{b}^{\|}}\right)\Pi(\mathbf{q}_{\|}),
\qquad\qquad\mbox{with}\qquad\qquad
\Pi(\mathbf{q}_{\|})=\frac{M_{\gamma}^{2}}{\mathbf{q}_{\|}^{2}},
\end{eqnarray}
with the photon mass
\begin{eqnarray}\label{CC-8-2}
M_{\gamma}^{2}=\frac{e^{2}N_{f}|eB|}{2\pi^{2}}.
\end{eqnarray}
According to the structure of $\Pi_{ab}(\mathbf{q}_{\|})$ in
(\ref{CC-8}), there are no polarization effects in the transverse
directions, and the strong screening effect appears only in the
longitudinal components of the photon propagator
$\sim\left(g^{ab}\mathbf{q}_{\|}^{2}-{\mathbf{q}^{a}_{\|}\mathbf{q}^{b}_{\|}}\right)$.
\par
In an alternative treatment, it is possible to work with the
ordinary gauge fields $A_{\mu}$ instead of the smeared fields
${\cal{A}}_{\mu}$. After redefining (\ref{CC-7}) in terms of
$A_{\mu}$, the vacuum polarization tensor is given by
\begin{eqnarray}\label{CC-8-1}
\widetilde{\Pi}_{ab}(\mathbf{q}_{\|},\mathbf{q}_{\perp})=\left(g_{ab}\mathbf{q}_{\|}^{2}-{\mathbf{q}_{a}^{\|}\mathbf{q}_{b}^{\|}}\right)
\widetilde{\Pi}(\mathbf{q}_{\|},\mathbf{q}_{\perp}),
\qquad\qquad\mbox{with}\qquad\qquad
\widetilde{\Pi}(\mathbf{q}_{\|},\mathbf{q}_{\perp})=\frac{M_{\gamma}^{2}(\mathbf{q}_{\perp})}{\mathbf{q}_{\|}^{2}},
\end{eqnarray}
and the ``effective mass''
$$
M_{\gamma}^{2}(\mathbf{q}_{\perp})=\frac{e^{2}N_{f}|eB|}{2\pi^{2}}e^{-\frac{\mathbf{q}^{2}_{\perp}}{2|eB|}},
$$
depending on the transverse coordinates. Keeping in mind that the
LLL approximation is only valid when $\mathbf{q}^{2}_{\perp}\ll
|eB|$, the phase factor $e^{-{\mathbf{q}^{2}_{\perp}}/{2|eB|}}$ can
be neglected, and we arrive at the same momentum independent mass
(\ref{CC-8-2}). This confirms the result computed in \cite{loskutov,
gusynin-1} using similar arguments.
\par
Now, in analogy to the evaluation of the anomaly in the previous
chapter, we will check the special role played by the momentum
$\mathbf{q}_{\perp}=\mathbf{0}$. To do this we compactify two
transverse coordinates around a circle with radius $R$. The
two-point vertex function (\ref{CC-5}) is therefore given by
\begin{eqnarray}\label{CC-9}
\Gamma_{LLL}^{(2)}=-\frac{ie^{2}N_{f}|eB|}{2\pi^{2}}\
\int\limits_{-\infty}^{+\infty}\frac{d^{2}q_{\|}}{(2\pi)^{2}}\
\frac{1}{(2R)^{2}}\sum\limits_{\mathbf{q}_{\perp}}
e^{-\frac{\mathbf{q}_{\perp}^{2}}{2|eB|}}A^{a}(q)
\left(g_{ab}-\frac{\mathbf{q}_{a}^{\|}\mathbf{q}_{b}^{\|}}{\mathbf{q}_{\|}^{2}}\right)
A^{b}(-q),
\end{eqnarray}
where we have used (\ref{YY-1}). Separating the zero and the nonzero
modes, we have first
\begin{eqnarray*}
\Gamma_{LLL}^{(2)}&=&-\frac{ie^{2}N_{f}|eB|}{2\pi^{2}}\
\int\limits_{-\infty}^{+\infty}\frac{d^{2}q_{\|}}{(2\pi)^{2}}\
\frac{1}{(2R)^{2}}\sum\limits_{\mathbf{q}_{\perp}\neq\mathbf{0}}
e^{-\frac{\mathbf{q}_{\perp}^{2}}{2|eB|}}\bar{A}^{a}(\mathbf{q}_{\|},\mathbf{q}_{\perp})
\left(g_{ab}-\frac{\mathbf{q}_{a}^{\|}\mathbf{q}_{b}^{\|}}{\mathbf{q}_{\|}^{2}}\right)
\bar{A}^{b}(-\mathbf{q}_{\|},-\mathbf{q}_{\perp})\nonumber\\
&&-\frac{ie^{2}N_{f}|eB|}{2\pi^{2}}\
\int\limits_{-\infty}^{+\infty}\frac{d^{2}q_{\|}}{(2\pi)^{2}}\
\frac{1}{(2R)^{2}}
A^{a,(0)}_{}(\mathbf{q}_{\|},\mathbf{q}_{\perp}=\mathbf{0})
\left(g_{ab}-\frac{\mathbf{q}_{a}^{\|}\mathbf{q}_{b}^{\|}}{\mathbf{q}_{\|}^{2}}\right)
A^{b,(0)}_{}(-\mathbf{q}_{\|},-\mathbf{q}_{\perp}=\mathbf{0}),
\end{eqnarray*}
where $\bar{A}$ denotes the nonzero and $A^{(0)}$ the zero
transverse modes of QED in a strong magnetic field. Using the
Fourier transformations (\ref{BB-28}) and (\ref{BB-29}) for the
gauge fields, we get
\begin{eqnarray*}
\Gamma_{LLL}^{(2)}&=&-\frac{ie^{2}N_{f}|eB|}{2\pi^{2}}\int\limits_{-\infty}^{+\infty}d^{2}y_{\|}\int\limits_{-R}^{+R}d^{2}y_{\perp}
e^{\frac{\nabla^{2}_{\perp}}{2|eB|}}\
\bar{A}^{a}_{}(\mathbf{y}_{\|},\mathbf{y}_{\perp})\left(g_{ab}-
\frac{\partial_{a}\partial_{b}}{\nabla_{\|}^{2}}\right)\bar{A}_{}^{b}(\mathbf{y}_{\|},\mathbf{y}_{\perp})\nonumber\\
&&-\frac{ie^{2}N_{f}|eB|}{2\pi^{2}}\int\limits_{-\infty}^{+\infty}d^{2}y_{\|}\frac{1}{(2R)^{2}}\int\limits_{-R}^{+R}d^{2}y_{\perp}A^{a,(0)}_{}(
\mathbf{y}_{\|},\mathbf{y}_{\perp})\left(g_{ab}-
\frac{\partial_{a}\partial_{b}}{\nabla_{\|}^{2}}\right)\int\limits_{-R}^{+R}d^{2}z_{\perp}A^{b,(0)}_{}(
\mathbf{y}_{\|},\mathbf{z}_{\perp}).
\end{eqnarray*}
Taking the limit $R\to\infty$, the second term vanishes and we are
left with
\begin{eqnarray}\label{CC-11-1}
\Gamma_{LLL}^{(2)}&=&-\frac{ie^{2}N_{f}|eB|}{2\pi^{2}}\int\limits_{-\infty}^{+\infty}d^{2}y_{\|}d^{2}y_{\perp}
e^{\frac{\nabla^{2}_{\perp}}{2|eB|}}\
\bar{A}^{a}_{}(\mathbf{y}_{\|},\mathbf{y}_{\perp})\left(g_{ab}-
\frac{\partial_{a}\partial_{b}}{\nabla_{\|}^{2}}\right)\bar{A}_{}^{b}(\mathbf{y}_{\|},\mathbf{y}_{\perp}).
\end{eqnarray}
Hence the zero transverse mode does not contribute to the vacuum
polarization tensor of the theory.
\section{Conclusions}
\noindent In this paper, we have calculated the $U_{A}(1)$ anomaly
of a $3+1$ dimensional QED in a strong and homogeneous magnetic
field in the lowest Landau level (LLL) approximation. Due to the
well established dimensional reduction $D\to D-2$ in the dynamics of
fermion pairing in a magnetic field, this anomaly is comparable with
the axial anomaly of a $1+1$ dimensional Schwinger model. On the
other hand, it is comparable with the axial anomaly of the ordinary
noncommutative field theory. The motivation behind this comparison
was the recently explored connection between the dynamics of
relativistic field theories in a strong magnetic field in the LLL
dominant regime and that in conventional noncommutative field
theories. Different aspects of the axial anomaly of noncommutative
$U(1)$ gauge theory are studied widely in the literature. Among
other results, it is well-known that noncommutative QED consists of
{\it two} anomalous axial vector current, whose anomalies receive
contribution from planar and nonplanar diagrams of the theory,
separately. Nonplanar (invariant) axial anomaly is affected by UV/IR
mixing, a phenomenon which is absent in the dynamics of a $3+1$
dimensional QED in a strong magnetic field.
\par
Apart from the fact that the axial anomaly of QED in the regime of
LLL dominance depends, in contrast to the ordinary $3+1$ dimensional
QED, linearly on the field strength tensor $F_{\mu\nu}$, it is
comparable with the {\it nonplanar} anomaly of noncommutative $U(1)$
gauge theory. To show this, we have compactified the transverse
directions to the external magnetic field along a circle with radius
$R$. A procedure which was also performed in the case of nonplanar
anomaly of noncommutative QED to explore the UV/IR mixing
phenomenon. We have shown that in the limit $R\to\infty$, the {\it
unintegrated} form of the axial anomaly of QED in the LLL
approximation receives contribution {\it only} from the nonzero
modes of the field strength tensor in the transverse directions. In
the case of nonplanar anomaly, however, the zero mode of the Fourier
transformed of ${\cal{F}}=F\widetilde{F}$ in the noncommutative
coordinates contributes to the {\it unintegrated} form of the
nonplanar anomaly only for finite $R$. In the limit $R\to\infty$,
the zero mode contribution and thus the unintegrated form of the
nonplanar anomaly vanish. We have further shown that the
contribution from the zero transverse mode of the field strength
tensor, {\it i.e.} the mode which is constant in the transverse
direction to the external magnetic field, reappears in the {\it
integrated} version of the axial anomaly of the QED in the LLL
approximation. The mechanism of this reappearance is quite similar
with the mechanism in which the {\it integrated} form of the
nonplanar anomaly was shown to be finite in $R\to\infty$ limit. The
main reason for all these effects, is the fact that QED in a strong
magnetic field consists of a natural UV cutoff $M\sim \sqrt{eB}$
which is kept large but finite, whereas noncommutative field
theories are treated as fundamental theories with infinitely large
UV cutoff.
\par
Further, motivated by the connection between the axial anomaly and
the vacuum polarization tensor in the ordinary two dimensional
Schwinger model, we have calculated the two-point vertex function of
QED in the LLL approximation. We have shown that the theory consists
of a massive photon of mass $M_{\gamma}\sim e^{2}|eB|$. This is in
analogy to the case of ordinary two dimensional Schwinger model
whose massive photon picks up a mass $m_{\gamma}^{2}=g^{2}/\pi$.
This can be again interpreted as a signature for the above mentioned
dimensional reduction from $D=4$ to $D=2$ dimensions.
\section{Acknowledgment}
\noindent The authors thank F. Ardalan for useful discussions. N.S.
thanks also S. Adler and A.-Y. Wang for their comments after
submitting the preprint to the hep-archive.

\end{document}